\documentclass[final,5p,times,twocolumn]{elsarticle}

\usepackage{amssymb}
\usepackage{amsthm}
\usepackage{amsmath}
\usepackage{siunitx}
\usepackage{ulem}
\usepackage{xcolor}
\usepackage{listings}
\usepackage{url}

\journal{~}
\graphicspath{{figures/}}

\begin{document}

\begin{frontmatter}



\title{A high-performance elliptic solver for plasma boundary turbulence codes}

\ead{Andreas.Stegmeir@ipp.mpg.de}
\author[label1]{Andreas Stegmeir}
\author[label2]{Cristian Lalescu}
\author[label2]{Mou Lin}
\author[label1]{Jordy Trilaksono}
\author[label3]{Nicola Varini}
\author[label2]{Tilman Dannert}

\address[label1]{Max Planck Institute for Plasma Physics, D-85748 Garching, Germany}
\address[label2]{Max Planck Computing and Data Facility (MPCDF), D-85748 Garching, Germany}
\address[label3]{Ecole Polytechnique F\'{e}d\'{e}rale de Lausanne (EPFL), SCITAS, CH-1015 Lausanne, Switzerland}

\begin{abstract}
Elliptic equations play a crucial role in turbulence models for magnetic confinement fusion. Regardless of the chosen modeling approach \--- whether gyrokinetic, gyrofluid, or drift-fluid \--- the Poisson equation and Amp\`{e}re's law lead to elliptic problems that must be solved on 2D planes perpendicular to the magnetic field. In this work, we present an efficient solver for such generalised elliptic problems, especially suited for the conditions in the boundary region. A finite difference discretisation is employed, and the solver is based on a flexible generalised minimal residual method (fGMRES) with a geometric multigrid preconditioner. We present implementations with OpenMP parallelisation and GPU acceleration, with backends in CUDA and HIP. On the node level, significant speed-ups are achieved with the GPU implementation, exceeding external library solutions such as \texttt{rocALUTION}. In accordance with theoretical scaling laws for multigrid methods, we observe linear scaling of the solver with problem size, $O(N)$. This solver is implemented in the \texttt{PARALLAX/PAccX} libraries and serves as a central component of the plasma boundary turbulence codes \texttt{GRILLIX} and \texttt{GENE-X}.

\end{abstract}


\end{frontmatter}

\section{Introduction}
Magnetic confinement fusion presents a promising, clean, and sustainable solution to meet the world’s growing energy demands, with ITER \cite{barabaschi:iter25} and SPARC \cite{creely:sparc20} serving as key milestones. Numerical simulations play a crucial role in interpreting experimental data and gaining a fundamental understanding of the complex physical mechanisms at play. As the performance of fusion reactors is primarily determined by turbulent processes, numerical tools are indispensable in the design process of future reactors and to advance toward the construction of fusion power plants.  In the core region, local gyrokinetic models \cite{brizard:gyrokinetics07} are typically the primary method. Research has increasingly focused on the boundary region, i.e.~the edge and scrape-off layer (SOL), to address key questions related to exhaust \cite{eich:prl11,chang:lambdaq17} and improved confinement regimes \cite{wagner:hmode07}. Due to higher collisionality, gyrofluid \cite{madsen:gyrofluid13} or drift-fluid \cite{zeiler:drbrag97} models can be viable alternatives. However, the boundary region presents significant complexity due to intricate magnetic geometry and strong fluctuation amplitudes.

All these turbulence models are based on the assumption of a strong and static background magnetic field $\mathbf{B}_0$, such that frequencies of interest are much smaller than the gyrofrequency $\omega \ll q_sB_0/m_s$ with $q_s$ and $m_s$ the charge and mass of particle species $s$. The turbulent structures are highly elongated along magnetic field lines $(k_\parallel \ll k_\perp)$, and their motion perpendicular to the magnetic field is described by drifts. One such drift, the $E\times B$ drift, arises from fluctuations in the electric field, which are generated intrinsically by microscopic plasma instabilities. In any plasma turbulence model, the electrostatic potential $\phi$ is obtained by solving a set of 2D generalised Poisson equations in planes perpendicular to the magnetic field (drift planes). Additionally, the treatment of electromagnetic perturbations $\tilde{\mathbf{B}}$ leads to similar equations for the parallel component of the perturbed magnetic vector potential $\tilde{A}_\parallel$. Throughout a turbulence simulation, this set of positive definite elliptic equations must be solved at each time step, potentially with coefficients that vary in both space and time. As such, an efficient solver is crucial for the performance of plasma turbulence codes.

In turbulence models for the core region, the assumption of small fluctuations ($\delta f$-approximation) and the use of globally field-aligned coordinate systems offer significant advantages. Spectral methods, combined with the Fast Fourier Transform (FFT), are highly efficient, and further discretization methods often yield linear systems of equations with sparse, time-invariant, band-diagonal matrices. Sparse direct solver libraries can handle these systems efficiently, with the costly matrix factorization only needing to be performed once at the beginning of the simulation \cite{germaschewski:genegpu21,latu:gyselasolver12}. Although in principle applicable for core turbulence models, the algorithms and code presented in this manuscript truly demonstrate their advantages when applied to elliptic problems arising in boundary turbulence models. The complex geometry in the boundary region, resulting from the presence of a separatrix and open field lines, makes the use of a globally field-aligned coordinate system impractical, thus eliminating the possibility of using Fourier methods. The presence of strong fluctuations requires the application of a full-f model, where the Boussinesq approximation \cite{ross:nonboussinesq18} is formally not justified, which leads to a generalised Poisson equation with spatio-temporally dependent coefficients. Consequently, the matrix associated with the discrete problem is no longer time-constant, and the costly factorisation step of direct solvers would have to be performed at each time step. 

In this manuscript, we introduce an efficient solver for generalised positive definite elliptic problems on 2D drift planes, designed specifically for boundary turbulence models. The equations are discretized using finite difference methods with a locally Cartesian mesh that remains logically unstructured. The solver employs an iterative approach based on the flexible generalised minimal residual method (fGMRES) \cite{saad:fgmres93}, leveraging the power of a geometric multigrid \cite{hackbusch:multigrid85} preconditioner to enhance performance significantly. The solver is implemented within the \texttt{PARALLAX} library \cite{phoenix-public:parallax25}, which has boosted edge turbulence simulations in both the drift-fluid code \texttt{GRILLIX} \cite{stegmeir:grillix19,zholobenko:hmode24} and the gyrokinetic code \texttt{GENE-X} \cite{michels:genex21}. The original solver was implemented in Fortran, utilising OpenMP for parallelisation. We have recently ported it to CUDA into a specialised library called \texttt{PAccX} (Parallax Accelerator) \cite{phoenix-public:paccx25}, enabling its compatibility with NVIDIA graphics processing units (GPUs). Through the application of the HIPifly \cite{amd:hipifly25} method, which requires only minimal modifications to the source code, we have expanded its vendor-portability to also include AMD GPUs.

Section \ref{sec:Field_equation_and_discretisation} provides the physical motivation and mathematical background, followed by the problem's discretisation using finite differences. Section \ref{sec:Algorithm_and_implementation} introduces the algorithm, with an emphasis on the geometric multigrid method. Section \ref{sec:implementation} details its implementation for CPU with OpenMP parallelisation and GPU architecture using CUDA and HIPifly. In Section \ref{sec:Benchmarks}, extensive benchmarks are conducted on the solvers, highlighting their linear scaling with problem size $O(N)$ and the significant speed-up achieved by the GPU implementation. We also depict the solver's application when integrated into production-scale turbulence simulations of \texttt{GRILLIX}. Finally, Section \ref{sec:Conclusions_and_outlook} concludes and projects future perspectives.

\section{Field equations and discretisation}
\label{sec:Field_equation_and_discretisation}

\subsection{Physical and mathematical background}
Tokamaks and stellarators confine plasma in toroidal geometry, using a strong background magnetic field. Typically, the toroidal magnetic field component is much stronger than the poloidal components $B_{tor}=R B^\varphi \gg B^R, B^Z$, where a cylindrical coordinate system $(R,\varphi, Z)$ is introduced with the central axis of the torus at $R=0$. In diverted reactors, inside the separatrix, the field lines spiral helically around the torus, forming nested flux surfaces. Outside the magnetic separatrix, in the scrape-off layer (SOL), the field lines end on target plates. 

The strong guide field leads to vastly different dynamics along (parallel to) and perpendicular to the magnetic field lines. While the charged plasma particles can stream freely along the magnetic field lines, their perpendicular motion follows a drift behavior, with the ExB drift $\mathbf{v}_{E\times B}=\mathbf{B}/B^2\times\nabla\phi$ one of the major drifts. However, the electrostatic potential $\phi$ is not externally imposed. Under the constraint of quasi-neutrality, the plasma must be regarded as a collective medium, where electric and magnetic field perturbations are generated intrinsically. The exact form of the field equations governing the electrostatic potential $\phi$ and the parallel component of the perturbed vector potential $\tilde{A}_\parallel$ depends on the specific model details. As an illustrative example, we discuss the field equation for $\phi$ in a gyro-fluid model, similarly to the one used in the \texttt{FELTOR} code \cite{wiesenberger:feltor24}, with a single ion species of charge $q_i$:
\begin{align}
-\nabla\cdot\left(\frac{m_iN_i}{B^2}\nabla_\perp\phi\right) = q_in_i-en_e,
\label{eq:phi_gyrofluid}
\end{align}
where $\nabla_\perp = \nabla -\mathbf{b}\mathbf{b}\cdot\nabla$ is the gradient perpendicular to the magnetic field with unit vector $\mathbf{b}$. On the right-hand side of this equation, the terms represent the charge density, with $n_e$ being the electron density and $n_i$ the ion density. However, in a gyro-fluid model, it is the ion gyro-center density $N_i$ that is evolved in time, rather than the ion density itself. These quantities are related through the gyro-average operator \cite{madsen:gyrofluid13, held:pade20}. Under the Pad\`e approximation, the ion density can be derived from the gyro-center density by solving an additional elliptic equation:
\begin{align}
\left(1-\nabla\cdot\frac{\rho_i}{2}\nabla_\perp\right)n_i = N_i,
\end{align}
where $\rho_i=\sqrt{T_{\perp,i}m_i}/(q_iB)$ is the thermal ion Larmor radius, with perpendicular ion temperature $T_{\perp,i}$. 

In full-f models, which are required to capture the strong fluctuations in the boundary region, the application of the Boussinesq approximation is formally not justified. As a result, a typical characteristic of full-f boundary models is the appearance of some form of density under the divergence in the field equation (\ref{eq:phi_gyrofluid}). Finally, we observe that the field equations in most boundary turbulence codes \--- such as \texttt{BOUT++} \cite{zhu:boutpp21}, \texttt{FELTOR} \cite{wiesenberger:feltor24}, \texttt{GBS} \cite{giacomin:gbs22}, \texttt{GENE-X} \cite{michels:genex21}, \texttt{GRILLIX} \cite{zholobenko:hmode24,zholobenko:grillixthermal20}, \texttt{GYSELA} \cite{grandgirard:gysela16}, and \texttt{XGC-1} \cite{ku:xgc09} \--- can be expressed as a generalised elliptic equation of the following form:
\begin{align}
\lambda \phi -\xi\nabla\cdot\left( c\nabla_\perp \phi\right) = \rho,
\label{eq:elliptic}
\end{align}
where $\lambda$, $\xi$ and $c$ are known real valued coefficients, potentially varying in space and time, and $\rho$ may represent some form of charge density. We note that we introduced the coefficient $\xi$ only for practical purposes, as it can be absorbed as well into $\lambda$ and $\rho$. So, for further theoretical considerations, we may assume $\xi=1$. Multiplying equation (\ref{eq:elliptic}) by $\phi^*$, integrating over the entire domain, and applying integration by parts \--- under the assumption that boundary terms vanish due to appropriately chosen boundary conditions \--- we obtain:
\begin{align}
\int\limits_V \phi^*\left(\lambda -\nabla\cdot\left( c\nabla_\perp \right)\right)\phi \, dV= \int\limits_V \lambda\left|\phi\right|^2 + c\left|\nabla_\perp\phi\right|^2\, dV.
\end{align}
The problem is positive definite if $c > 0$ and $\lambda\geq 0$, which is the class of problem relevant for boundary turbulence models, where $c$ typically represents some form of density.

The boundary region is also characterized by complex geometry due to the presence of both open and closed field lines and the separatrix, which introduces singularities in global field-aligned coordinates. Consequently, discretization often relies on non-aligned or Flux Coordinate Independent (FCI) \cite{hariri:fci13,stegmeir:fci16} approaches, where the mesh consists of a set of poloidal planes at toroidal angles $\varphi_k$ with toroidal mesh index $k$. In the general coordinate representation, the differential operator in Eq.~(\ref{eq:elliptic}) is expressed as:
\begin{align}
\nabla\cdot\left(c\nabla_\perp\phi\right) = \frac{1}{\sqrt{g}}\partial_i\left[c\sqrt{g}\left(g^{ij}-b^ib^j\right)\right]\partial_j,
\end{align}
where Einstein summation convention is applied, and $g^{ij}$ is the metric tensor. Assuming that the toroidal magnetic field is much stronger than the poloidal magnetic field ($b^\varphi\sim 1$), the problem reduces to a set of two-dimensional problems within poloidal planes. If we adopt a cylindrical coordinate system and absorb the Jacobian into the coefficients ($c\rightarrow cR$, $\xi\rightarrow\xi/R$), the elliptic problem \ref{eq:elliptic} becomes:
\begin{align}
\lambda\phi - \xi\partial_R\left(c\partial_R\phi\right) - \xi\partial_Z\left(c\partial_Z\phi\right) = \rho.
\label{eq:elliptic_coordinate}
\end{align}
The elliptic problem is well-posed only when appropriate boundary conditions are specified. We assume that the boundary of the domain $\Omega$ is divided into disjoint subsets, $\partial\Omega=\partial\Omega_D \cup \partial\Omega_N$, where either Dirichlet or Neumann boundary conditions are applied:
\begin{align}
\phi = \phi_D \quad\text{on}\quad \partial\Omega_D, \label{eq:dirichlet_boundary_condition} \\
\mathbf{n}\cdot\nabla\phi = \Gamma \quad\text{on}\quad \partial\Omega_N, \label{eq:neumann_boundary_condition}
\end{align}
where $\mathbf{n}$  denotes the unit normal vector on the Neumann boundary $\partial\Omega_N$.

\subsection{Discretisation}
\begin{figure}
    \includegraphics[trim=0 0 0 0, clip=true, width=0.45\textwidth]{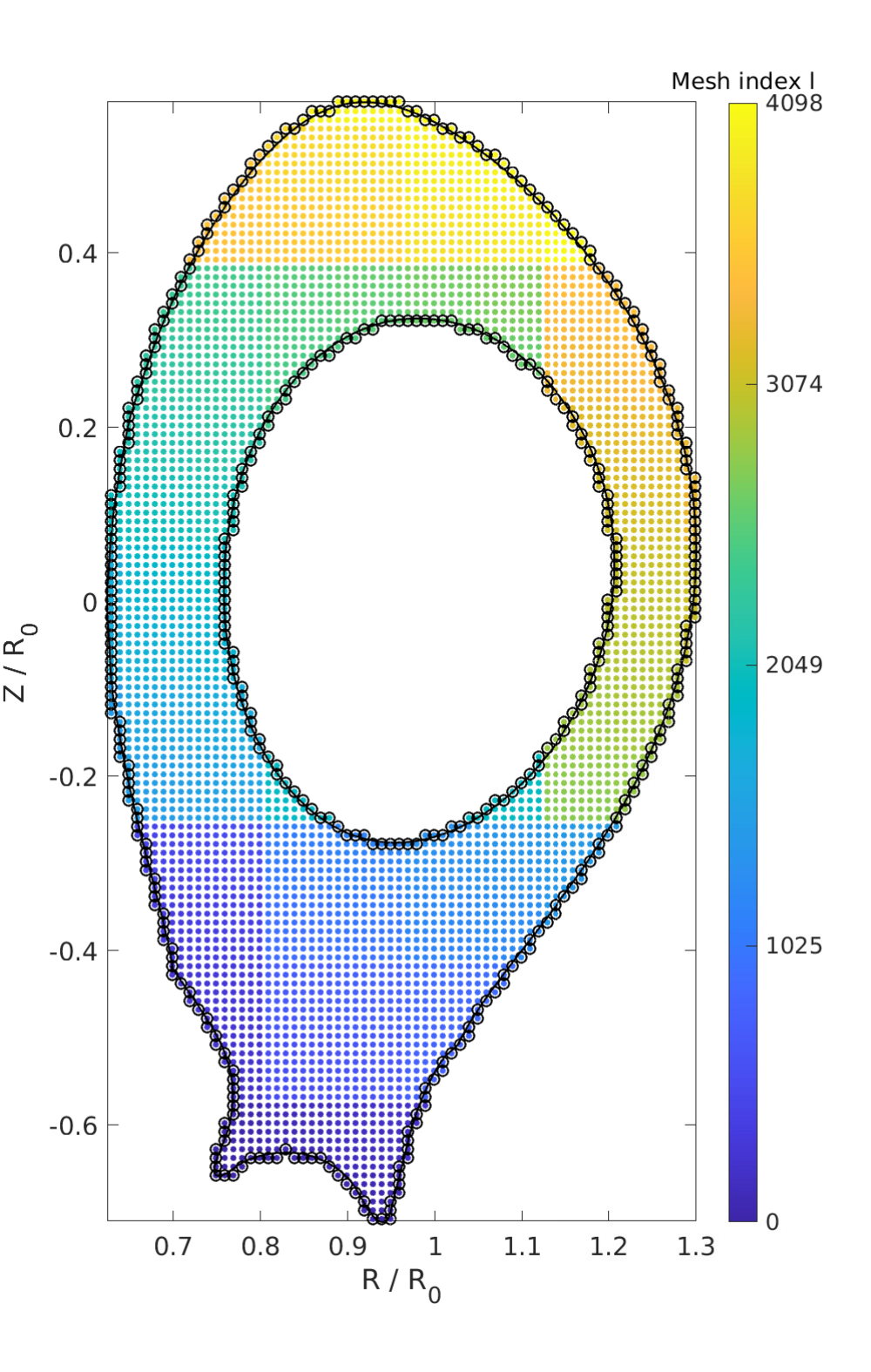}
    \includegraphics[trim=0 0 0 0, clip=true, width=0.45\textwidth]{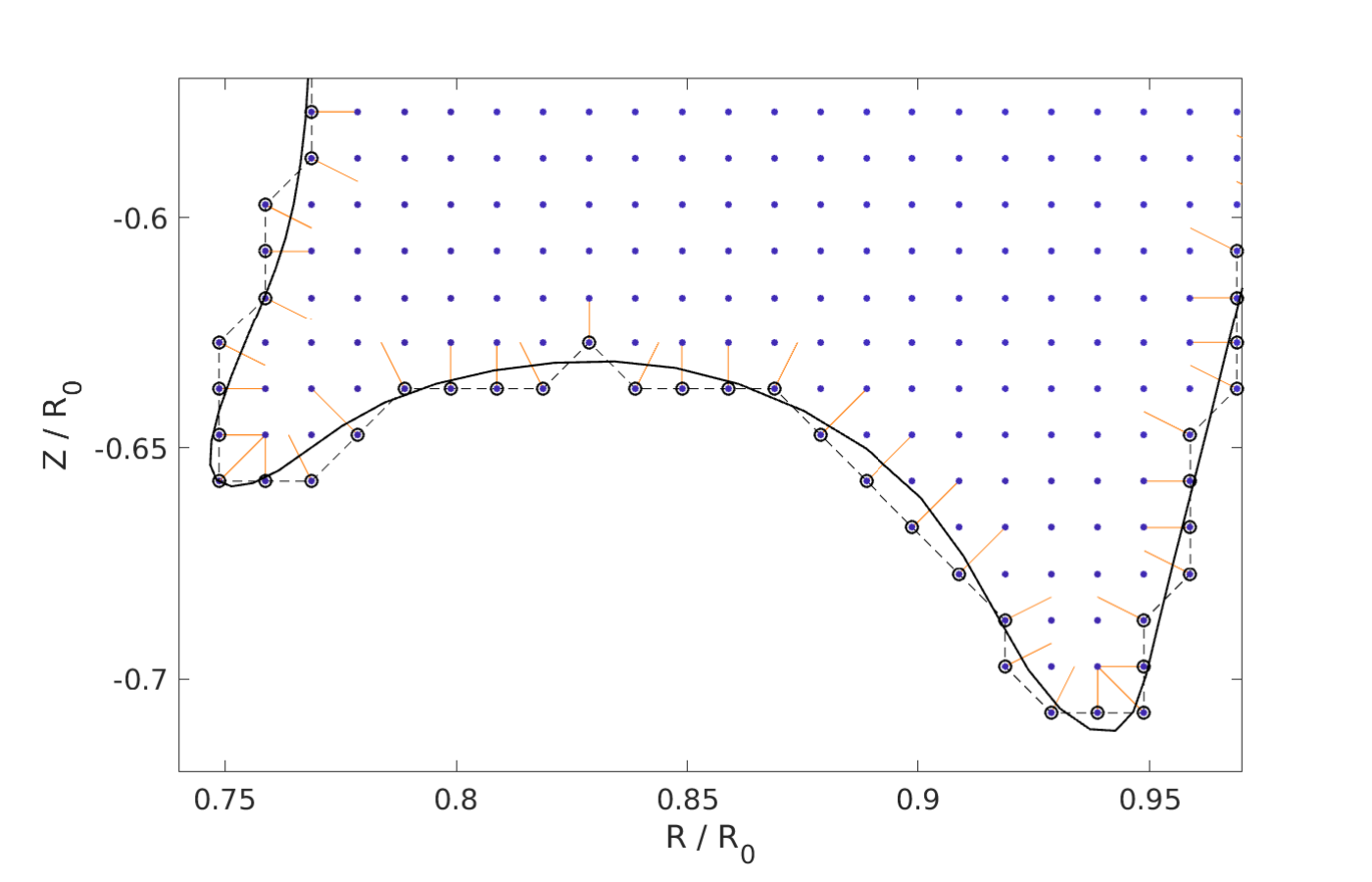}
    \caption{Top: Example of a mesh representing a single poloidal plane in a typical tokamak boundary simulation setup. The color indicates the mesh index $l$, which can be reordered arbitrarily. The inner computational points are surrounded by guard points (black circles), which wrap the continuously prescribed boundary (black line). \\
    Bottom: A zoomed-in view of the divertor region, showing the handling of boundary conditions. The dashed black line represents the discrete boundary contour, and the orange lines show the discrete normals to the boundary, from which upstream values $\phi_N$ are obtained to compute the Neumann boundary conditions.}
    \label{fig:mesh2d}
\end{figure}

We introduce a logically unstructured mesh that is locally Cartesian (see Fig.~\ref{fig:mesh2d}). The coordinates of the mesh points are given by:
\begin{align}
    R_l = & R_{ref} + i_l h, & Z_l = & Z_{ref}+ j_l h,
\end{align}
where ($R_{ref}$, $Z_{ref}$) denotes a reference point and $h$ is the grid spacing. The integer lists $i_l$ and $j_l$ represent the discrete horizontal and vertical positions for a given grid point with mesh index $l=1\dots N$. The following methods and algorithms are independent of the mesh ordering, allowing for flexibility in choosing different mesh orderings to optimize cache performance. The mesh consists of inner points, each of which has both vertical and horizontal neighbors, and guarding boundary points that lack at least one neighboring point and serve to approximate the continuous boundary.
 
At each inner grid point, we discretize the elliptic problem on a single plane (\ref{eq:elliptic_coordinate}) using a straightforward second-order finite difference scheme, resulting in a five-point stencil operation:

\begin{align}
\lambda_{i,j}\phi_{i,j} - \frac{\xi_{i,j}}{2}[&\left(c_{i+1,j} + c_{i,j}\right)\phi_{i+1,j} + \left(c_{i,j} + c_{i-1,j}\right)\phi_{i-1,j} \notag \\
+ & \left(c_{i,j+1} + c_{i,j}\right)\phi_{i,j+1} + \left(c_{i,j-1} + c_{i,j}\right)\phi_{i,j-1} \notag \\
- & \left(c_{i+1,j}+c_{i-1,j} + c_{i,j+1} + c_{i,j-1} +4c_{i,j}\right)\phi_{i,j} ] = \rho_{i,j}.
\label{eq:discrete_fieldeq}
\end{align}
For brevity, we used here a notation with Cartesian indices $i$ and $j$, noting that the mapping to the unstructured mesh index $(i,j)\rightarrow l$ is available. For boundary points that are part of a Dirichlet boundary we simply set
\begin{align}
\phi_l = \phi_{D,l}, \quad \text{if} \quad (R_l, Z_l) \in \partial\Omega_D^h,
\label{eq:discrete_dirichlet}
\end{align}
where $\partial\Omega_D^h$ is the set of boundary points, where the Dirichlet boundary condition shall be applied. We note that the prescribed value may vary among the boundary points. 

For the discretization of the Neumann boundary conditions, we introduce a discrete level set function:
\begin{align}
\psi_{i,j}=
\begin{cases}
1 \quad \text{on inner mesh points} \\
\frac{1}{2} \quad \text{on boundary point} \\
0 \quad \text{if not part of the mesh}
\end{cases}
\end{align}
We then define the components of a discrete boundary normal for each boundary point (see Fig.~\ref{fig:mesh2d}, bottom):
\begin{align}
n^R_{i,j} = & \psi_{i+1,j}-\psi_{i-1,j}, & n^Z_{i,j} = &\psi_{i,j+1}-\psi_{i,j-1},
\end{align}
where we are again using Cartesian notation. The discrete normal components can take values in $\lbrace-1,\,-1/2,\, 0,\, 1/2,\, 1\rbrace$. The upstream value is then determined as:
\begin{align}
\phi_{i,j}^N = \phi_{i+n^R_{i,j},j+n^Z_{i,j}},
\end{align}
if both $n^R_{i,j}$ and $n^Z_{i,j}$are integer values. If, for example, $n^R_{i,j}$ is an integer and $n^Z_{i,j}$ is a half-integer, the upstream value is computed as:
\begin{align}
\phi_{i,j}^N = \frac{1}{2}\phi_{i+n^R_{i,j},j} + \frac{1}{2}\phi_{i+n^R_{i,j},j+2n^Z_{i,j}},
\end{align}
with analogous expressions for the further cases. The discrete version of the Neumann boundary condition (\ref{eq:neumann_boundary_condition}) is then
\begin{align}
\frac{\phi^N_{i,j} - \phi_{i,j}}{h\sqrt{\left(n^{R}_{i,j}\right)^2+\left(n^{Z}_{i,j}\right)^2}} = \Gamma_{i,j}, \quad \text{if} \quad (R_l, Z_l) \in \partial\Omega_N^h,
\label{eq:discrete_neumann}
\end{align}
where $\partial\Omega_N^h$ is the set of boundary points, where the Neumann boundary condition shall be applied. While the discretization of the inner points is second-order accurate, the treatment of the boundary conditions is expected to reduce the overall convergence rate. This is due to two factors: First, there is a mismatch between the continuous $(\partial\Omega)$ and discrete boundary contours $(\partial\Omega^h)$, with the latter only approximating the former in a stepwise manner. Second, the discretization of the Neumann boundary condition is somewhat ad hoc and less accurate. However, such a less precise treatment of the boundaries is generally acceptable in practical applications, where artificial buffer zones are often introduced to smooth the solution within these zones. Furthermore, a property of the boundary condition expressions, which will be useful in the following, is that each boundary point can be set independently, relying only on the prescribed boundary value and nearby interior mesh points, without depending on other boundary points.

\section{Algorithm}
\label{sec:Algorithm_and_implementation}

Equations (\ref{eq:discrete_fieldeq}, \ref{eq:discrete_dirichlet}) and (\ref{eq:discrete_neumann}) represent the discrete version of the elliptic problem (\ref{eq:elliptic_coordinate}, \ref{eq:dirichlet_boundary_condition}) and (\ref{eq:neumann_boundary_condition}), and can be expressed as a linear equation system:
\begin{align}
\mathbf{A}\mathbf{x}=\mathbf{b},
\label{eq:lineqsys}
\end{align}
where $\mathbf{x}$ contains the values of the solution $\phi_l$, and $\mathbf{b}$ the charge density $\rho_l$ at inner mesh points, as well as the prescribed boundary values $\phi_{D,l}$ and fluxes $\Gamma_l$ on boundary points. The matrix $\mathbf{A}$ is sparse with five non-zero entries per row corresponding to inner mesh points, and up to three entries per row for boundary points. In time-dependent simulations, the values of the entries may vary due to their dependency on the coefficients; however, the sparsity pattern remains fixed. The matrix is generally unstructured, but its structure can be optimized during the initialization phase of a simulation by adjusting the mesh ordering.

In the following, we present several approaches to solving this problem, including the use of external libraries. Given the structure of the matrix $\mathbf{A}$ and the intended application context, an abstract design pattern is proposed, consisting of three accessible routines:
\begin{itemize} \label{HAUBA}
    \item \texttt{create}: A factory routine that constructs and returns a \texttt{solver} object based on the specified algorithm and solver parameters. It initialises data structures and allocates the required memory. 
    \item \texttt{update}: Updates the internal data of the \texttt{solver} object with the new instance of coefficients or boundary conditions.
    \item \texttt{solve}: Computes the solution $\mathbf{x}$ for a given right-hand side $\mathbf{b}$ and initial guess $\mathbf{x}_0$.
\end{itemize}
Only the latter two routines are invoked within the time loop, making them performance critical.

\subsection{Multigrid preconditioner}
The fact that the matrix entries may vary between successive time steps motivates the use of iterative methods. We employ the flexible GMRES (Generalised Minimal Residual) algorithm introduced in~\cite{saad:fgmres03}. GMRES methods are generally robust and suited for solving systems with nonsymmetric matrices. However, their efficiency heavily depends on the availability of an effective preconditioner. The flexible variant of GMRES accommodates the use of varying, potentially iterative, preconditioners across Krylov iterations. For elliptic problems, multigrid methods are commonly used as preconditioners due to their rapid convergence rates that are independent of the grid size. Moreover, their computational cost scales linearly with the number of unknowns~\cite{hackbusch:multigrid85}. In the following, we present a geometric multigrid algorithm that includes the formulation of the discrete problem, Eqs.~(\ref{eq:discrete_fieldeq}, \ref{eq:discrete_dirichlet}) and (\ref{eq:discrete_neumann}) on coarsened mesh levels, transfer operations between successive levels, an appropriate smoother, and a solver for the coarsest level.

\subsubsection{Grid hierarchy and transfer operations}
\begin{figure}
    \includegraphics[trim=0 0 0 0, clip=true, width=0.5\textwidth]{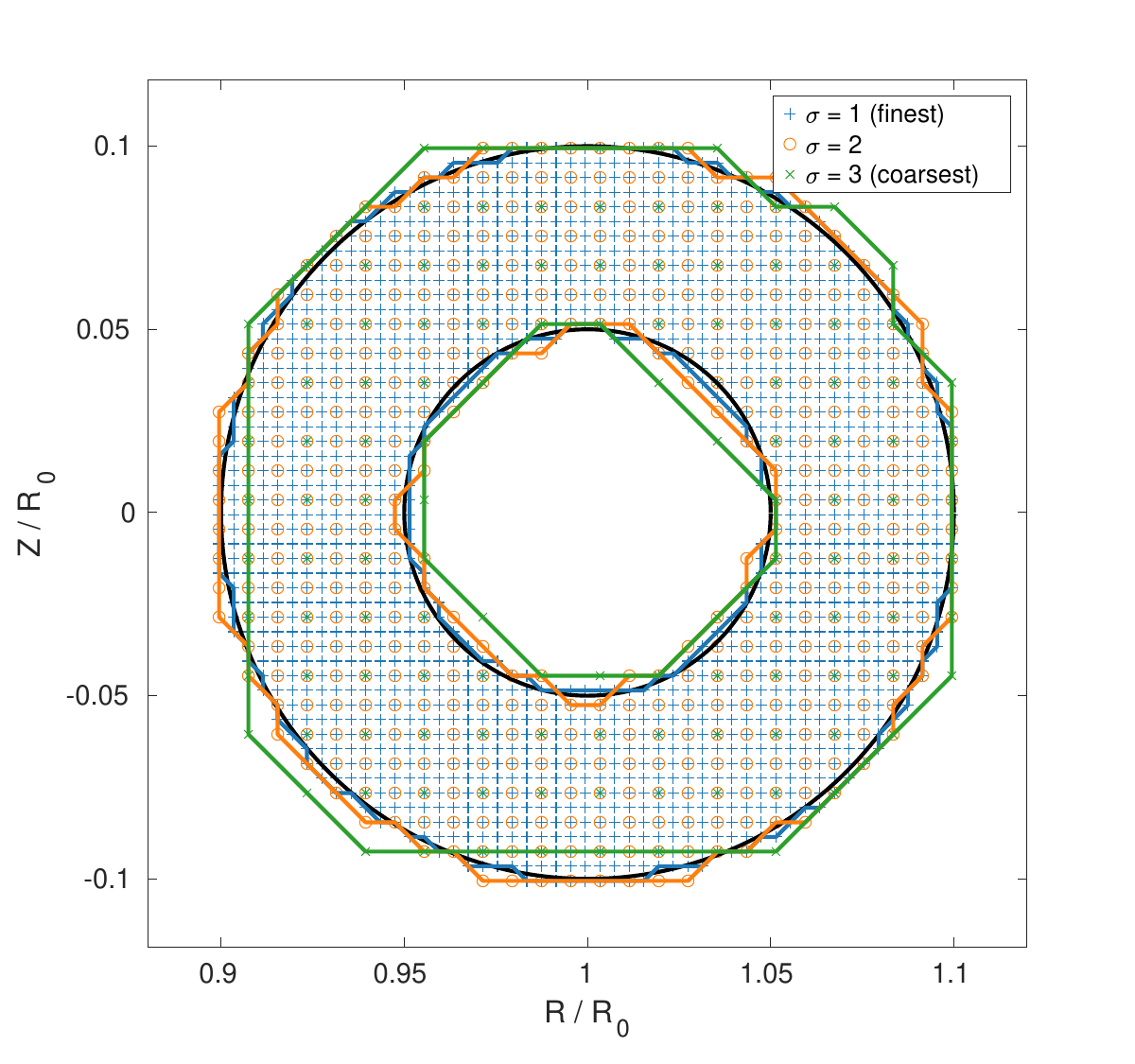}
    \caption{Example of a multigrid mesh hierarchy with three levels of coarseness $\sigma$. The discrete boundaries at each level are shown as colored solid lines, each passing through the corresponding mesh’s boundary points. These discrete boundaries approximate the continuous boundary (solid black line) as closely as possible.}
    \label{fig:multi_grids}
\end{figure}

We denote the coarseness level by $\sigma = 1,\dots, N_\sigma$, where $\sigma=1$ corresponds to the finest mesh level. Each coarser level mesh is constructed from every other point from the mesh at the next finer level, with the Cartesian resolution at each level $h_\sigma = h / 2^{\sigma-1}$. The meshes are then adjusted such that the boundary points approximate the continuous boundary as closely as possible. As illustrated in Fig.~\ref{fig:multi_grids}, the discrete boundary varies across mesh levels. This approach preserves the equidistant spacing of the meshes at each level and minimizes over- or under-constraining during mesh transfer operations caused by missing information. Otherwise, the discretization expressions on the coarser levels are derived straightforwardly from those on the finest level, with appropriate adjustments to the mesh spacing and indexing.

\begin{figure}
    \includegraphics[trim=0 0 0 0, clip=true, width=0.2\textwidth]{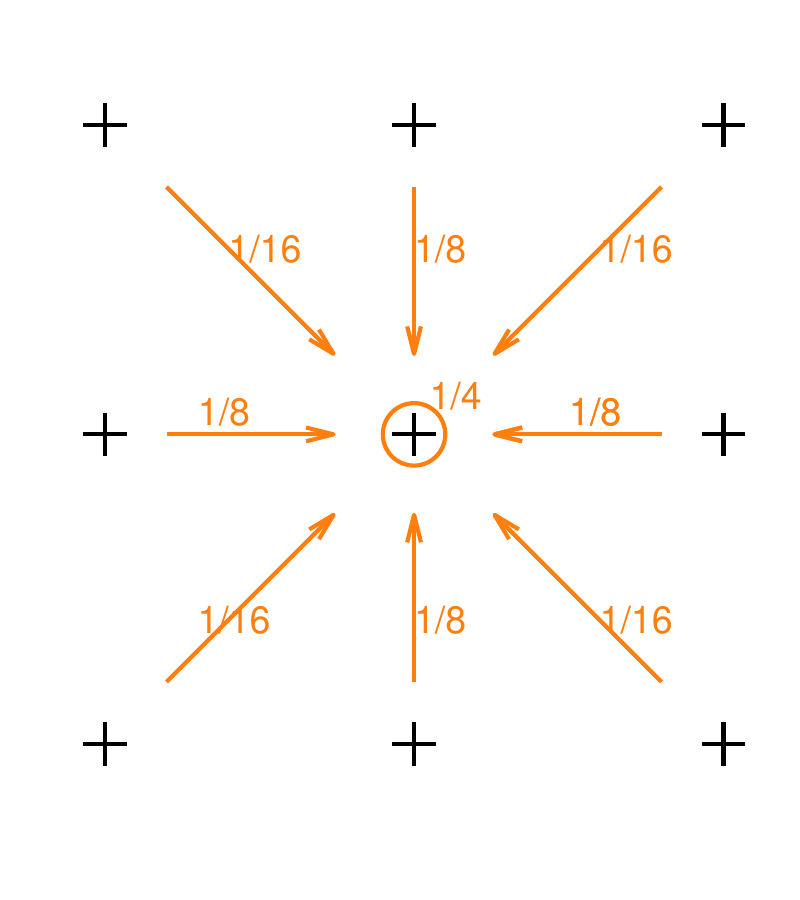} \hspace{0.5cm}
    \includegraphics[trim=0 0 0 0, clip=true, width=0.2\textwidth]{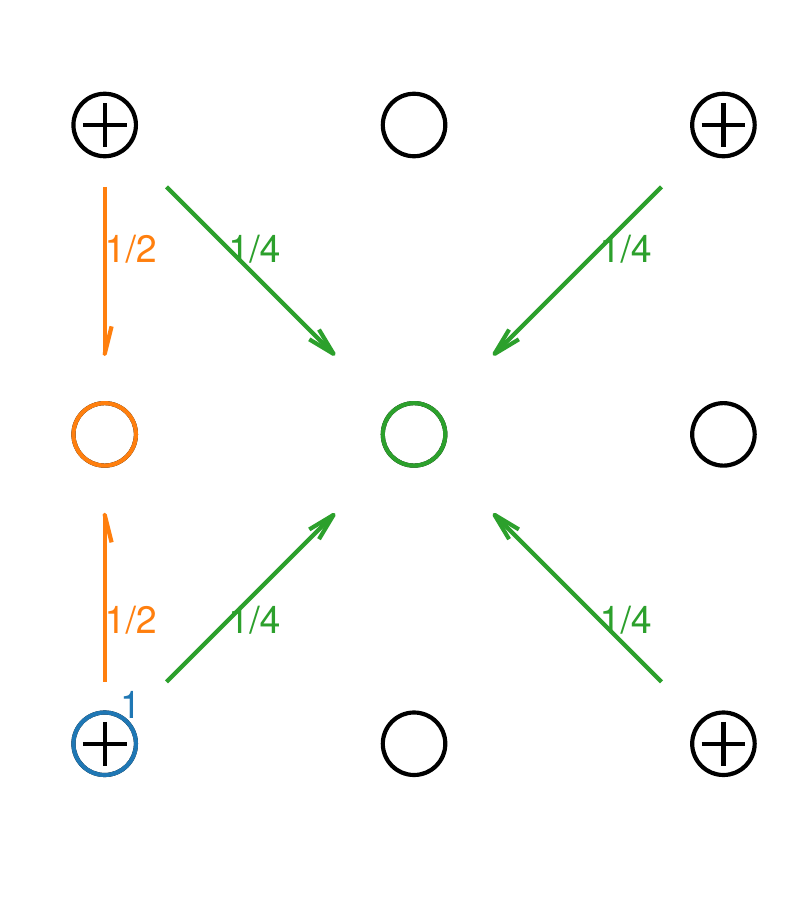}
    \caption{Left: Transfer to a coarse mesh point (orange circle) is achieved through a weighted restriction from surrounding fine mesh points (black crosses). Right: Transfer to fine mesh points (circles) from a coarser mesh (black crosses) is carried out using bilinear interpolation. The three cases, each resulting in different interpolation weights, are indicated with colors.}
    \label{fig:op_transfer}
\end{figure}

For restriction to the respective next coarser level $(\sigma\rightarrow\sigma + 1)$, a weighted averaging scheme is employed, and for prolongation to the respective next finer level $(\sigma\rightarrow\sigma - 1)$, bilinear interpolation is used (see Fig.~\ref{fig:op_transfer}). During transfer operations to boundary points, stencil points may be missing. Boundary values can explicitly be reset after each grid transfer based on the prescribed boundary conditions. Finally, we note that the coefficients $\lambda$, $\xi$, and $c$ undergo grid transfer operations as well.

\subsubsection{Smoother and coarsest level solver}
The Gauss-Seidel method is well-known as an efficient smoother in multigrid applications. However, with lexicographic ordering, the Gauss-Seidel smoother cannot be parallelised. To overcome this limitation, the mesh points on each level are divided into three independent subgroups \cite{bruaset:parallelnumerics06}. The interior mesh points are organized in a chequerboard pattern (red-black), with the third group consisting of the boundary points. These three groups are then updated in subsequent parallel subloops, according to the Gauss-Seidel algorithm. Due to the five-point stencil, the first parallel loop over red points does not depend on any other red point, but only on black and boundary points. Similarly, the second parallel loop over black points does not depend on any other black point. The discrete boundary expressions, Eqs.~(\ref{eq:discrete_dirichlet}) and (\ref{eq:discrete_neumann}), imply that the final parallel loop over boundary points does not depend on any other boundary point, but only on red and black points.

On the coarsest level, we use a sparse direct solver based on LU factorization, leveraging external libraries such as the \texttt{PARDISO} solver from MKL \cite{schenk:pardiso00} or \texttt{SuperLU} \cite{li:superlu18}. Although this factorization needs to be performed at each time step, it remains computationally efficient due to the low dimensionality of $\mathbf{A}$ on the coarsest level.

\subsubsection{Multigrid algorithm} \label{sec:multigrid_algorithm}
We follow the multigrid algorithm in a recursive form as outlined in \cite{meister:numlin15}:

\begin{quotation}
\noindent 
\texttt{recursive subroutine mg\_cycle(}$\mathbf{x}^\sigma$, $\mathbf{b}^\sigma$, $\sigma$\texttt{)}\newline
\indent \texttt{if (}$\sigma$ \texttt{==} $N_\sigma$\texttt{) then} \newline
\indent \indent \texttt{! Direct solve on coarsest level} \newline
\indent \indent $\mathbf{x}^\sigma = \left(\mathbf{A}^\sigma\right)^{-1}\mathbf{b}^\sigma$ \newline
\indent \indent \texttt{return} \newline
\indent \texttt{endif} \newline
\indent \texttt{! Presmoothing} \newline
\indent \texttt{do i = 1, nsmooth\_pre} \newline
\indent \indent $\mathbf{x}^\sigma$ \texttt{ = gauss\_seidel\_rbb(}$\mathbf{x}^\sigma,\mathbf{b}^\sigma$\texttt{)} \newline
\indent \texttt{enddo} \newline
\indent \texttt{! Restrict defect} \newline
\indent $\mathbf{d}^\sigma = \mathbf{A}^\sigma\mathbf{x}^\sigma - \mathbf{b}^\sigma$ \newline
\indent $\mathbf{d}^{\sigma+1} = \mathbf{R}^{\sigma\rightarrow\sigma+1}\mathbf{d}^\sigma$ \newline
\indent\texttt{! Compute correction }\newline
\indent $\mathbf{e}^{\sigma+1}_0= 0$\newline
\indent \texttt{do k = 1,}$\gamma$ \newline
\indent \indent $\mathbf{e}^{\sigma+1}_k$ \texttt{= mg\_cycle(}$\mathbf{e}^{\sigma+1}_{k-1},\mathbf{d}^{\sigma+1}$\texttt{)} \newline
\indent \texttt{enddo} \newline
\indent \texttt{! Prolong and apply correction} \newline
\indent $\mathbf{e}^{\sigma}_\gamma = \mathbf{P}^{\sigma+1\rightarrow\sigma}\mathbf{e^{\sigma+1}_{\gamma}}$ \newline
\indent $\mathbf{e}^\sigma_\gamma$ \texttt{ = set\_bnds\_hom(}$\mathbf{e}^\sigma_\gamma$\texttt{)} \newline
\indent $\mathbf{x}^\sigma = \mathbf{x}^\sigma - \mathbf{e}^\sigma_\gamma$ \newline
\indent \texttt{! Postsmoothing} \newline
\indent \texttt{do i = 1, nsmooth\_post} \newline
\indent \indent $\mathbf{x}^\sigma$ \texttt{ = gauss\_seidel\_rbb(}$\mathbf{x}^\sigma,\mathbf{b}^\sigma$\texttt{)} \newline
\indent \texttt{enddo} \newline
\texttt{end subroutine}
\end{quotation}

In the above pseudocode $\mathbf{A}^\sigma$ denotes the matrix of the discrete problem on level $\sigma$, $\mathbf{R}^{\sigma\rightarrow\sigma+1}$ the restriction matrix to the next coarser level $\sigma+1$ and $\mathbf{P}^{\sigma+1\rightarrow\sigma}$ is the prolongation matrix back to level $\sigma$. The number of pre- and post-smoothing steps is controlled via the parameters \texttt{nsmooth\_pre} and \texttt{nmsooth\_post} respectively. The type of multigrid cycle is determined via the parameter $\gamma$, where $\gamma=1$ results in a V-cycle and $\gamma=2$ in a W-cycle \cite{meister:numlin15}. A specialty of our implementation is the explicit setting of boundary conditions on the correction $\mathbf{e}_\gamma^\sigma$ after prolongation. As indicated by the function \texttt{set\_bnds\_hom}, we apply homogeneous Dirichlet values and Neumann fluxes on boundary points according to Eqs.~(\ref{eq:discrete_dirichlet}) and (\ref{eq:discrete_neumann}).

\section{Implementation}
\label{sec:implementation}

\subsection{\texttt{PARALLAX}'s field solver}

The field solver is part of the \texttt{PARALLAX} library, which also provides tools for processing tokamak equilibria and stellarator magnetic configurations, including field-line tracing capabilities essential for the FCI approach. The edge/SOL fluid turbulence code \texttt{GRILLIX} and the gyrokinetic code \texttt{GENE-X} are both built on the \texttt{PARALLAX} library, which provides the common numerical and computational foundation for handling their geometry, mesh, FCI operators, and field equations. \texttt{PARALLAX} is written in modern Fortran 2008, and we outline its components relevant to the field solver in the following.

The mesh object stores the positions of mesh points along with their connectivity information. Meshes can be independently generated at various levels of coarseness \--- constructing a coarser mesh does not require access to a finer one. The multigrid object then stores the meshes across all levels, along with the associated restriction and prolongation matrices that transfer quantities between successive levels. So far, the data is static \---  it is initialised once at the beginning of a simulation and remains unchanged between time steps. A dedicated routine is available to construct the discrete field matrix $\mathbf{A}^\sigma$ on any given grid level in Compressed Sparse Row (CSR) format, based on the coefficients $\lambda$, $\xi$, and $c$, as well as the specified type of boundary conditions.

The multigrid preconditioner object stores the matrices $\mathbf{A}^\sigma$ at each level, for which memory is allocated and the non-zero structure is defined during creation. As this information remains unchanged throughout the simulation, it is called before entering the time-step loop. An associated update routine, called at every time step, first restricts the coefficients to all coarser levels, from which the matrix entries for each $\mathbf{A}^\sigma$ are then derived. Additionally, the matrix on the coarsest level is LU-factorised. Finally, the cycle routine executes the multigrid algorithm as described in Section~\ref{sec:multigrid_algorithm}. The cycle routine is called at the finest level by the flexible GMRES solver, which is implemented using a reverse communication interface \cite{intel:mkl25}. A central driver routine manages the control flow by issuing tasks such as performing a matrix-vector multiplication or applying the preconditioner. After each task is completed, control returns to the driver routine. This process continues iteratively until a stopping criterion, a threshold on the residuum, is satisfied.

\texttt{PARALLAX} employs OpenMP parallelisation within poloidal planes, meaning that any loop iterating over the grid index (at any level) is parallelised. An additional MPI-parallelisation is introduced only in the application level, i.e.,\texttt{GRILLIX} and \texttt{GENE-X}, which involves domain decomposition along the toroidal angle and, in the case of \texttt{GENE-X}, additionally over velocity space and plasma species. Specifically, each poloidal plane can be assigned to a separate MPI process.

\subsection{GPU backends \-- PAccX library}
The few available GPU programming models for Fortran, such as OpenACC and CUDA Fortran, are not very mature and lack flexibility, often facing limitations in compiler support, especially for Fortran 2008. Our initial attempts to offload \texttt{PARALLAX} to GPUs using compiler-based directives failed early on, we believe, primarily due to compatibility issues of the compilers with the modern Fortran coding style. Therefore, we went for a mixed language approach using the intrinsic ISO\_C\_BINDING module, where relevant parts of \texttt{PARALLAX} are rewritten in C/C++, which offers more mature and flexible GPU programming models, such as CUDA and HIP. Instead of directly integrating the C/C++ source code into \texttt{PARALLAX}, we developed \texttt{PAccX} \--- the \texttt{PARALLAX} Accelerator Library \--- as a separate library exposing a pure C interface for the \texttt{PARALLAX} library. This separation simplifies code management and maintenance, as the development of the C/C++ GPU backend in \texttt{PAccX} is decoupled from the ongoing development of \texttt{PARALLAX}, enabling independent development cycles for both.

Static data \--- such as the multigrid object that stores meshes and transfer matrices \--- is still created within \texttt{PARALLAX} and exposed to Fortran-C interoperable structures for \texttt{PAccX} via pointers at the start of a simulation. The dynamic components, including the flexible GMRES solver, matrix assembly routines, and multigrid preconditioner, have been rewritten in C++. An abstract design pattern enables the selection of different backends for these routines, with a CUDA implementation currently available. Compatibility with AMD architectures is achieved using the HIPifly method \cite{amd:hipifly25}, which provides aliases for CUDA routines in HIP. This approach requires only minimal changes to the source code, mainly limited to minor adjustments in CUDA-related header files using preprocessor directives. Memory transfers between host and device occur during the initialisation phase, where essential static data is copied to the device. During the time loop, transfers are limited to the coefficients in the update stage and to the right-hand side, initial guess, and final solution in the solve stage. An exception to this is the direct solve on the coarsest level, which currently still relies on CPU-based libraries such as \texttt{PARDISO} or \texttt{SuperLU}. These will be replaced in the near future by GPU-based libraries or an in-house developed GPU direct solver. We note that \texttt{PARALLAX}/\texttt{PAccX} supports interfacing with data that resides already on the device. In a forward-looking manner, this eliminates memory transfers between host and device within the time loop entirely. However, since the other components of \texttt{GRILLIX} have not yet been ported to GPUs, this feature is not yet exploited, but will be implemented once the entire \texttt{GRILLIX} codebase supports GPU acceleration.


\subsection{External library solvers}
Thanks to the abstract design of the solver object, both \texttt{PARALLAX} and \texttt{PAccX} allow for straightforward integration of external solver libraries without interfering the APIs to applications. The matrix on the finest level, provided in CSR format, can be directly passed to an external library. This makes it particularly convenient to experiment with or benchmark against established solvers and preconditioners from libraries, particularly algebraic multigrid methods. We conducted preliminary tests with \texttt{PETSc} \cite{argonne:petsc25}, which offers a broad selection of solvers and preconditioners, including the \texttt{hypre} multigrid preconditioner \cite{hypre}. For our application, we found \texttt{rocALUTION} \cite{rocm:rocalution25} to be a promising library, and used its BiCGStab linear solver \cite{van1992bi} with smooth-aggregated algebraic multigrid (SAAMG) \cite{vanek1996algebraic} preconditioning. 
When applied directly to the discretized system \eqref{eq:lineqsys}, particularly in cases with Neumann boundary conditions, we first applied \texttt{rocALUTION} to a Jacobi preconditioner:
\begin{align}
\mathbf{D}^{-1}\mathbf{A}\mathbf{x}=\mathbf{D}^{-1}\mathbf{b},
\end{align}
with $\mathbf{D}$ the diagonal of the matrix $\mathbf{A}$. When solving this preconditioned system, we ensured that the residual norm still met the desired threshold relative to the original unpreconditioned system.
Since \texttt{rocALUTION} is a highly modular library, the application of these numerical solvers is very straightforward:
\begin{verbatim}
    BiCGStab<LocalMatrix<double>, 
             LocalVector<double>, 
                         double> ls_;
    SAAMG<LocalMatrix<double>, 
          LocalVector<double>, 
                    double> p_;
    ls_.SetOperator(mat_);
    ls_.SetPreconditioner(p_);
    ls_.Init(rel_tol_, 0.0, 1e+8, 10000);  
    ls_.Build();
    ls_.Solve(rhs_, &x_);
\end{verbatim}
One could also replace it with other linear solvers or preconditioners supported by \texttt{rocALUTION}.

\section{Benchmarks}
\label{sec:Benchmarks}

\subsection{Definition of test cases}
To verify and benchmark the implementation we consider a geometry with circular flux surfaces. For this setup, we introduce a poloidal flux label and a poloidal angle defined by:
\begin{align}
r = &\frac{\sqrt{(R-R_0)^2 + Z^2}}{R_0},
&\theta = &\arctan\frac{Z}{R-R_0}.
\end{align}
We consider a single poloidal plane, where the domain is bounded by an inner and outer flux surface located at $r_{\text{min}} = 0.2$ and $r_{\text{max}} = 0.4$, respectively. From this, we define the normalised flux label:
\begin{align}
\hat{r} = &\frac{r-r_{min}}{r_{max}-r_{min}}.
\end{align}
The following coefficient are used:
\begin{align}
\lambda = &r\sin\theta, \quad \xi = R\sqrt{r}\left(2+\cos\theta\right), \notag \\
c = &\frac{1}{R}\left[1.1+\cos\left(\frac{\pi}{2}\hat{r}\right)\cos\left(3\theta\right)\right].
\end{align}
In the spirit of the method of manufactured solutions \cite{salari:mms00}, we prescribe the following analytical solution:
\begin{align}
\phi_{\text{MMS}} = 1.3 + \cos\left(\frac{3}{2}\pi \hat{r}\right)\sin(4\theta),
\end{align}
from which the corresponding `charge density` $\rho$ can be computed analytically using equation~(\ref{eq:elliptic_coordinate}). Dirichlet boundary conditions are applied at the outer boundary, with values specified at locations of the boundary points. At the inner boundary, we consider either Dirichlet conditions or Neumann boundary conditions, for which the normal derivatives can also be obtained analytically.

This setup is general, as neither the boundaries nor the mesh are aligned with the functions defined in the test case. As such, it is representative of more complex geometries. The tests presented in the following have also been carried out in a similar manner for more intricate configuration \--- including diverted geometries with X-point(s) \--- and have demonstrated consistent behavior.

For the following tests, we began with a resolution of $h=4\cdot10^{-3}$, corresponding to  $N = 25,726$ grid points, where we employed $N_\sigma=4$ multigrid levels. We then progressively refined the grid by halving $h$ at each step and incrementing $N_\sigma$ by one. This process continued until we reached the finest resolution of $h=1.25\cdot10^{-4}$, yielding  $N = 24,196,346$ grid points, and where we employed $N_\sigma=9$ multigrid levels. This resolution range covers and goes well beyond the mesh sizes employed in typical production runs of \texttt{GRILLIX} and \texttt{GENE-X}, which are between $N\approx1\cdot10^5$ and $N\approx2\cdot10^{6}$.

While \texttt{PAccX} supports V-, W-, and F-cycles, we focus exclusively on benchmarking the V-cycle, as it consistently demonstrated the best overall performance on GPUs. Unless otherwise specified, five pre- and post-smoothing steps are used, and \texttt{SuperLU} serves as the direct solver on the coarsest level. The stopping criterion for the GMRES algorithm is based on the normalised residual,
\begin{align}
\text{res} = \frac{\left| \mathbf{Ax} - \mathbf{b} \right|}{\left| \mathbf{b} \right|},
\end{align}
which is required to fall below a threshold of $\text{res}<1\cdot10^{-8}$.

\subsection{Verification}
To verify the correctness of the implementation, we solve the system using the analytically prescribed charge density, coefficients, and boundary conditions, and compare the resulting numerical solution to the analytical MMS solution. The numerical error is evaluated using the normalised L2 norm and the supremum norm. We conduct the analysis for two different cases: (1) Dirichlet boundary conditions at both the inner and outer boundaries, and (2) Neumann boundary conditions at the inner boundary with Dirichlet conditions at the outer boundary.

\begin{figure}
    \includegraphics[trim=0 0 0 0, clip=true, width=0.5\textwidth]{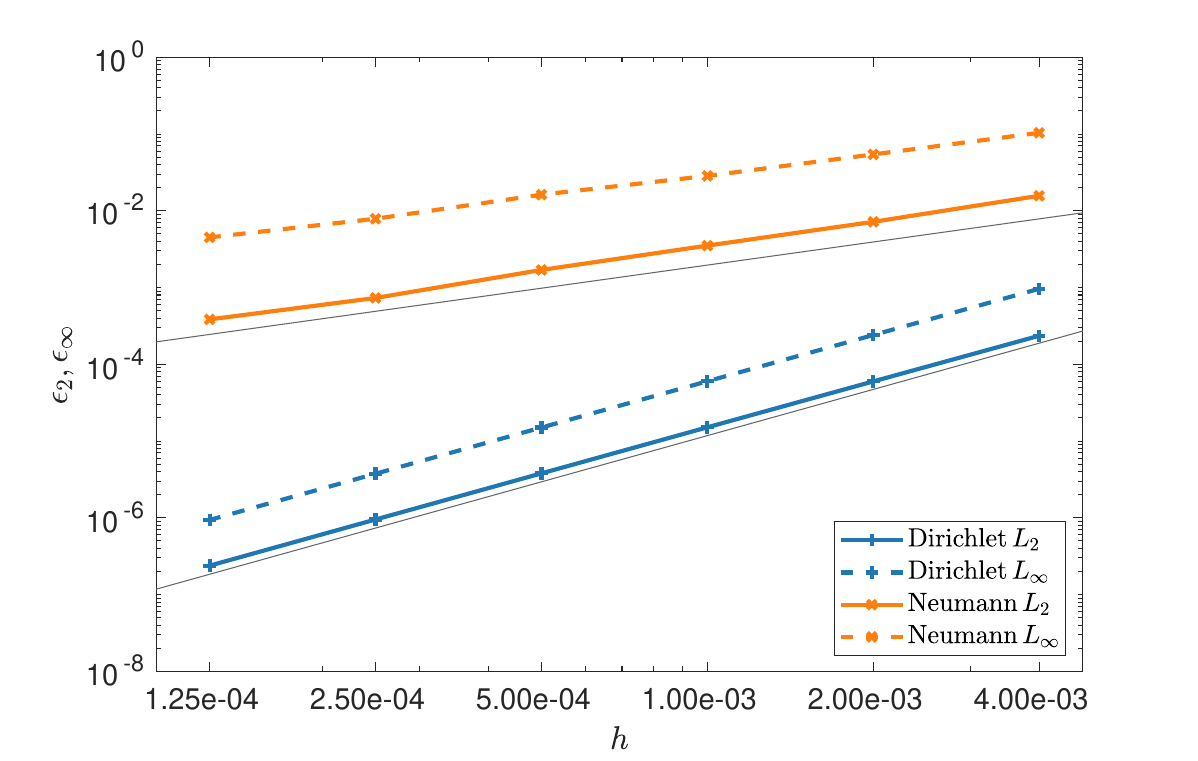}
    \caption{Numerical error of the field solver in the $L_2$ norm (solid) and supremum norm $L_\infty$ (dashed) for the case with Dirichlet (blue) and Neumann (orange) boundary conditions at the inner boundary. The black reference lines indicate first-order ($\propto h$) and second-order ($\propto h^2$) convergence rates.}
    \label{fig:convergence_field_solver}
\end{figure}

As shown in Fig.~\ref{fig:convergence_field_solver}, the case with Dirichlet boundary conditions exhibits second-order convergence, consistent with expectations based on the discretization scheme. The case with Neumann boundary conditions achieves only first-order convergence, which can be attributed to the lower-order approximation used for computing the normal fluxes at the boundary, as described in equation (\ref{eq:discrete_neumann}). This verification was performed across all available backends (Fortran, CXX, CUDA, HIP) and on multiple computing systems. The results were consistent across all configurations, agreeing to within less than $1\cdot10^{-8}$. The number of iterations required for convergence was identical across all backends.

\subsection{Scaling with problem size}
We benchmark our application on the Raven supercomputer \cite{mpcdf:raven25} at the Max Planck Computing and Data Facility (MPCDF). Its GPU nodes feature Intel Xeon IceLake-SP processors with 72 cores and 4 Nvidia A100 GPUs. For a fair assessment of the GPU acceleration, we compare the CPU-based Fortran implementation with OpenMP parallelisation running on a quarter of the node (18 cores) vs.~the CUDA backend from \texttt{PAccX} running on one GPU. It is worth noting that the OpenMP version shows no significant performance improvement when using more than 18 cores. We measure the execution time for the update and the solve stages of the solvers separately, which are the performance-relevant sections in applications.

\begin{figure}
    \includegraphics[trim=0 0 0 0, clip=true, width=0.5\textwidth]{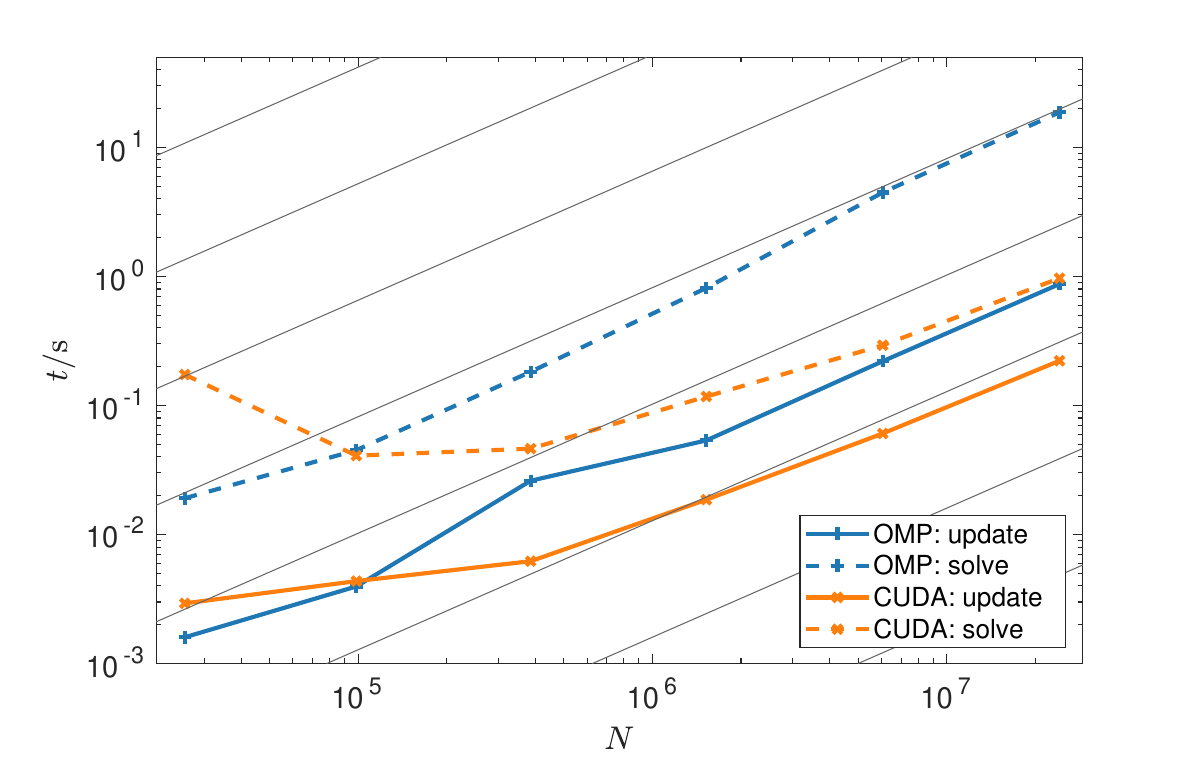}
    \includegraphics[trim=0 0 0 0, clip=true, width=0.5\textwidth]{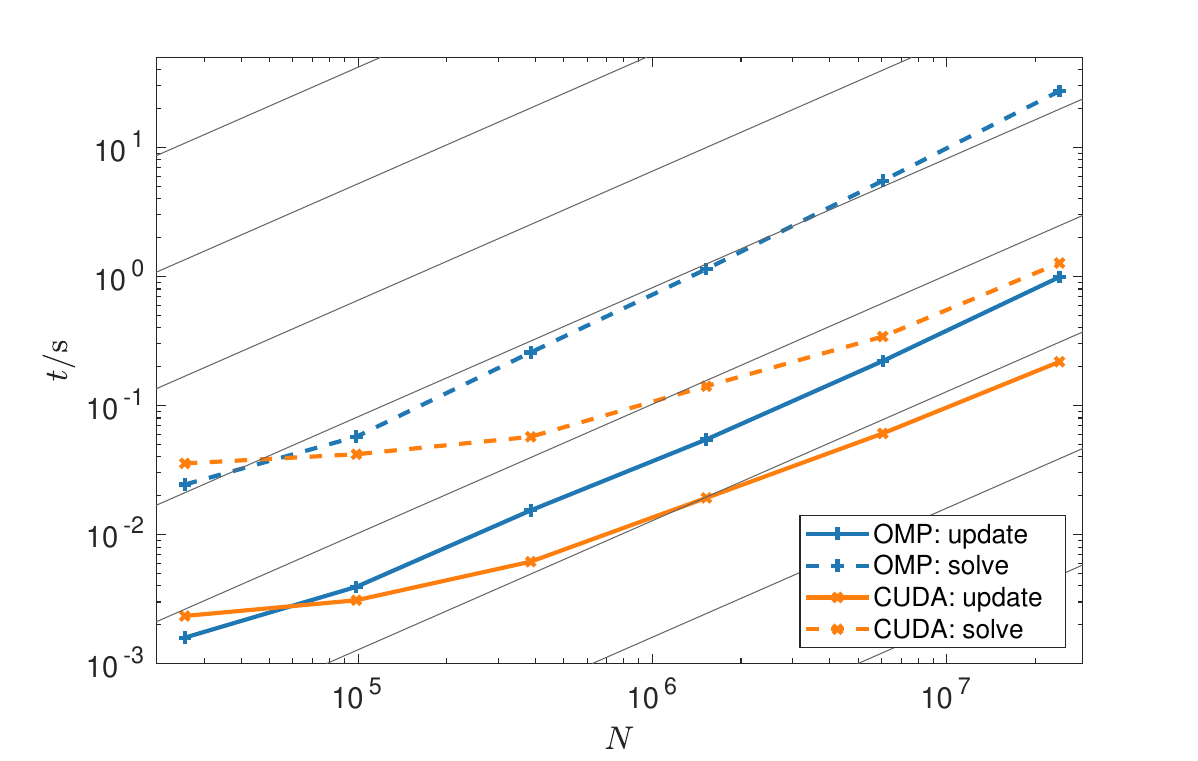}
    \caption{Benchmark of the CPU-based solver with OpenMP parallelisation (blue) versus the GPU-accelerated solver using the CUDA backend (orange), for a test case with Dirichlet (top) and Neumann (bottom) boundary conditions at the inner boundary. Execution times are shown separately for the update (dashed) and the solve stage (solid). Both exhibit linear scaling with the problem size, $O(N)$, where $N$ denotes the number of grid points. Gray reference lines indicate ideal linear scaling.}
    \label{fig:benchmark_nscaling}
\end{figure}

Fig.~\ref{fig:benchmark_nscaling} presents the scaling of execution time with problem size for different backends, considering both Neumann and Dirichlet boundary conditions at the inner boundary.  Due to the use of a sparse matrix format, the update routine is expected to scale linearly with problem size, i.e., $O(N)$. Theory on multigrid methods predicts linear scaling for the solve stage as well \cite{hackbusch:multigrid85}, as the convergence speed does not deteriorate when the discretisation is refined. Across the wide range of resolutions tested, the number of GMRES iterations required to reduce the residuum below the threshold increases from 11 to 17. The linear scaling with problem size is eventually recovered in the CPU-based implementation using OpenMP across a wide range of problem sizes. For the GPU-accelerated implementation, the linear scaling shows up only at largest problem sizes. At the largest problem size considered, the GPU implementation is a factor of $\approx15$ faster than the CPU implementation.

\subsection{Contrasting different backends and machines}
\begin{figure}
    \includegraphics[trim=0 0 0 0, clip=true, width=0.5\textwidth]{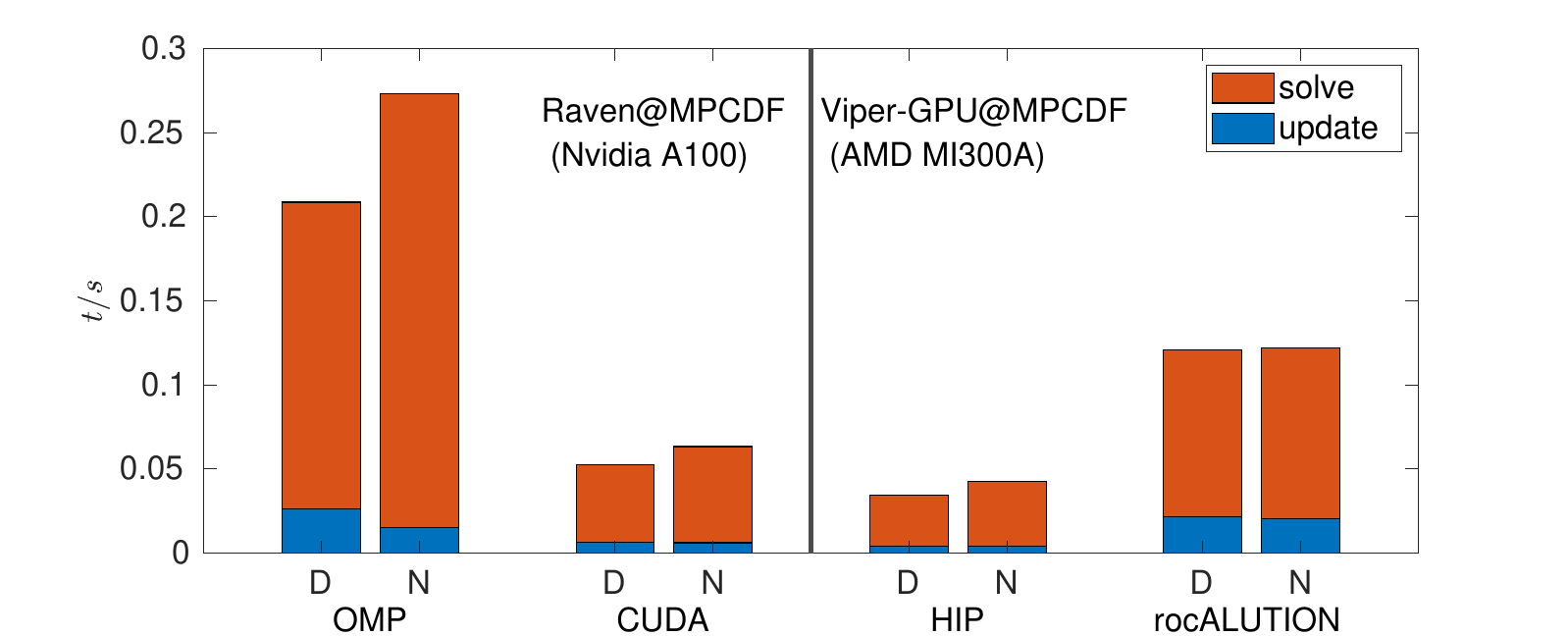}
    \includegraphics[trim=0 0 0 0, clip=true, width=0.5\textwidth]{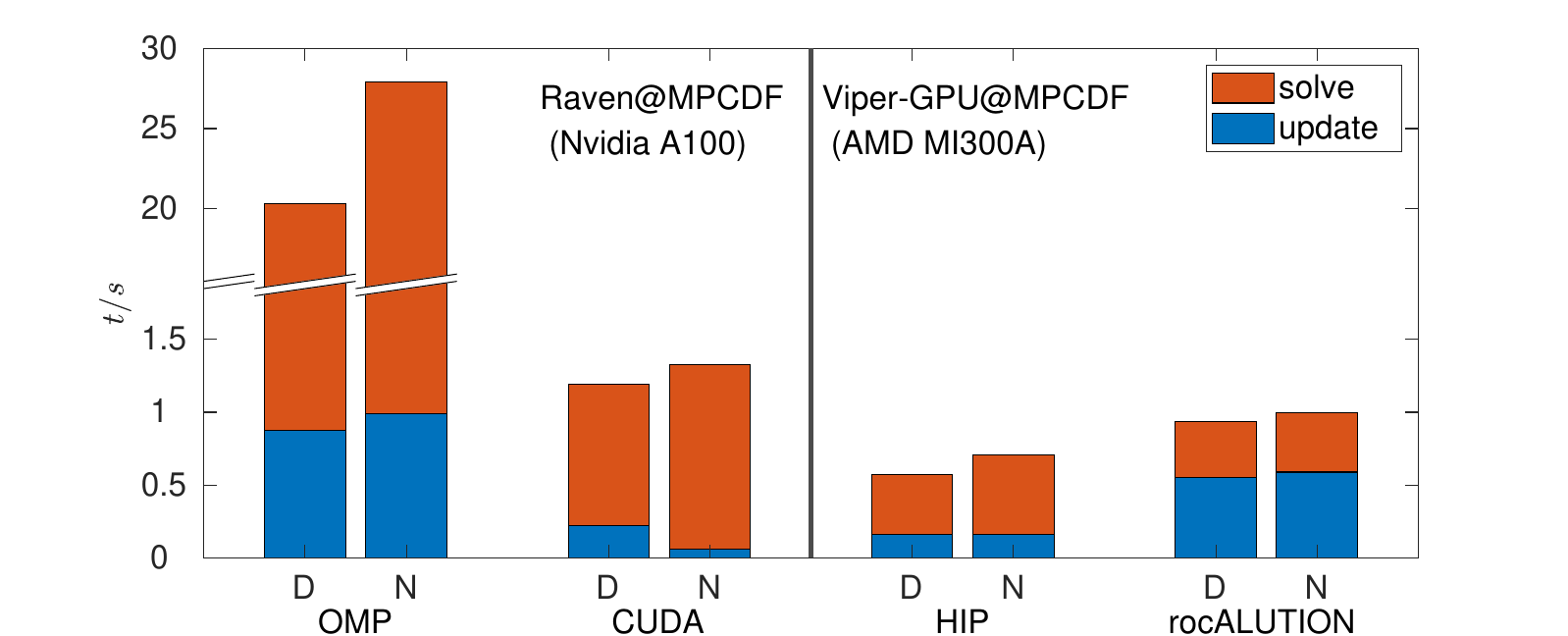}
    \caption{Execution time of the solver using different backends for a medium-sized problem with $N = 385,611$ points (top) and a large problem with $N = 24,196,346$ points (bottom). The OpenMP and CUDA backends were executed on the Raven system equipped with an NVIDIA A100 GPU (left half of each plot), while the HIP and rocALUTION backends were run on the Viper system with an AMD MI300A GPU (right half). For each backend, the left bar represents execution time with Dirichlet boundary conditions (D), and the right bar with Neumann boundary conditions (N).}
    \label{fig:backends_all}
\end{figure}

Figure~\ref{fig:backends_all} presents a performance comparison using different computational backends. The OpenMP and CUDA backends were executed on the Raven supercomputer at MPCDF, as described in the previous section. The HIP and \texttt{rocALUTION} backends were evaluated on the Viper-GPU supercomputer, also hosted by MPCDF, which features two AMD Instinct MI300A APUs per node. Results are shown for two problem sizes: a medium-sized problem, representative of current production runs with \texttt{GRILLIX} and \texttt{GENE-X}, and a large-scale problem, representative of near-future, reactor-relevant applications. 

The results clearly demonstrate a substantial performance gain when using our CUDA-based implementation of the geometric multigrid preconditioner, compared to the OpenMP-based Fortran version. For the medium-sized case, we observe an approximate $4\times$ speedup on the GPU, increasing to more than $15\times$ for the large-scale case. We observe that the HIP backend running on the MI300A is very performant as  well, achieving roughly twice the speed of the CUDA backend on the Nvidia A100. This performance difference is consistent with the respective hardware specifications \cite{nvidia:a100datasheet25, amd:MI300Adatasheet25}. In this context, we also compare our custom implementation to \texttt{rocALUTION}, a library-based solution. Our solver appears a bit faster than \texttt{rocALUTION}, particularly for smaller problem sizes, while also offering greater portability. Nevertheless, the results show that \texttt{rocALUTION} can be a competitive alternative, and we note that further fine-tuning of algorithmic parameters \--- such as the mesh reordering strategy described in the following section \--- remains possible for all backends, which may lead to slight variations in the reported performance numbers. However, when using such library-based solvers \--- including experiments with algebraic multigrid methods provided by \texttt{PETSc} \--- we encountered occasional robustness issues, where certain problem configurations failed to converge below the desired residuum. In contrast, our geometric multigrid algorithm has consistently demonstrated high robustness across all tested scenarios, a result validated through its extensive and stable use in both \texttt{GRILLIX} and \texttt{GENE-X}.

\subsection{Performance breakdown}
\label{sec:performance_breakdown}
\begin{figure*}
    \includegraphics[trim=0 0 0 0, clip=true, width=1.0\textwidth]{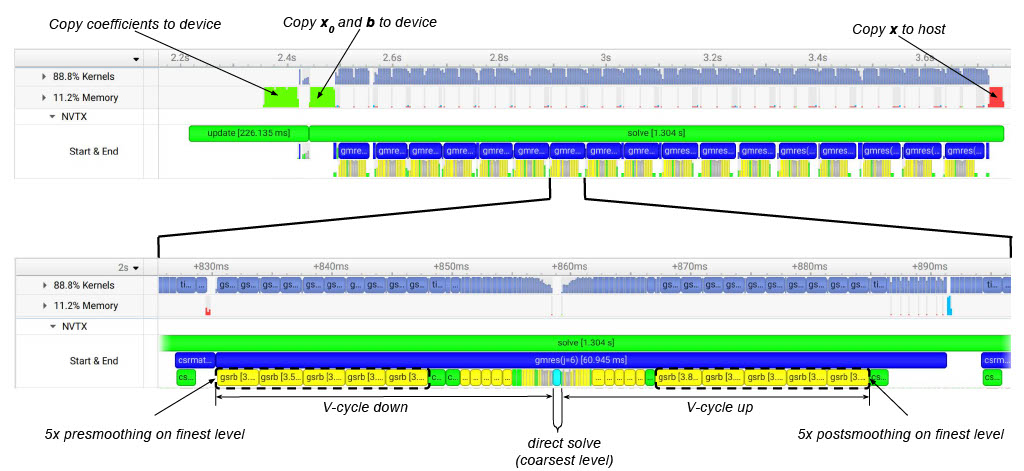}
    \caption{Excerpt from the NVIDIA Nsight Systems performance analysis tool for the large case $(N = 24,196,346)$ with Neumann boundary conditions. The top section shows the timeline of the complete update and solve phase, while the bottom section provides a zoomed-in view of a single multigrid V-cycle.}
    \label{fig:nsight_paccx}
\end{figure*}

To obtain a detailed breakdown of the solver performance, we profile the large case with Neumann boundary conditions using the NVIDIA Nsight Systems tool \cite{nvidia:nsight25} on Raven@MPCDF, employing the CUDA backend (see Fig.~\ref{fig:nsight_paccx}). Memory transfer operations \--- primarily coefficient data during the update phase and the initial guess, right-hand side, and solution vectors during the solve phase \--- account for $11.2\%$ of the total execution time. The remaining $89.8\%$ is spent in GPU compute kernels. The most significant contributions stem from smoothing operations. Smoothing on the finest level already accounts for approximately $50\%$ of the compute time, whereas grid transfer operations and smoothing on all coarser levels combined account for about $30\%$. The direct solver requires a very small fraction of compute time in this case, and the remainder is spent on further operations related to the GMRES algorithm. The distribution of computing time depends on system size, where, towards smaller cases, the share of the direct solve, which is still performed on the CPU, starts to become significant.

\subsection{Effect of mesh reordering}
The flexible mesh ordering offers an opportunity for fine-tuning performance. Initially, the mesh is constructed using a lexicographic ordering, with indices assigned row-wise from the lower-left to the upper-right mesh point. Still during the initialisation phase, the mesh can be arbitrarily reordered to optimize cache efficiency. Morton ordering \cite{morton:zorder66} (or Z-order) preserves spatial locality by storing nearby points in memory locations that are close to each other. This ordering can be extended by grouping neighboring elements into small contiguous blocks (e.g., $8\times8$) and applying Morton ordering to these blocks. Such block-based Morton ordering can be particularly beneficial for finite difference methods with compact stencils, as used in \texttt{GRILLIX} and \texttt{GENE-X}, since it enhances cache efficiency by improving data locality. In contrast, the red-black Gauss-Seidel algorithm used in the geometric multigrid method benefits from a red-black ordering of the mesh, which aligns better with its alternating update pattern.

\begin{figure}
    \includegraphics[trim=0 0 0 0, clip=true, width=0.5\textwidth]{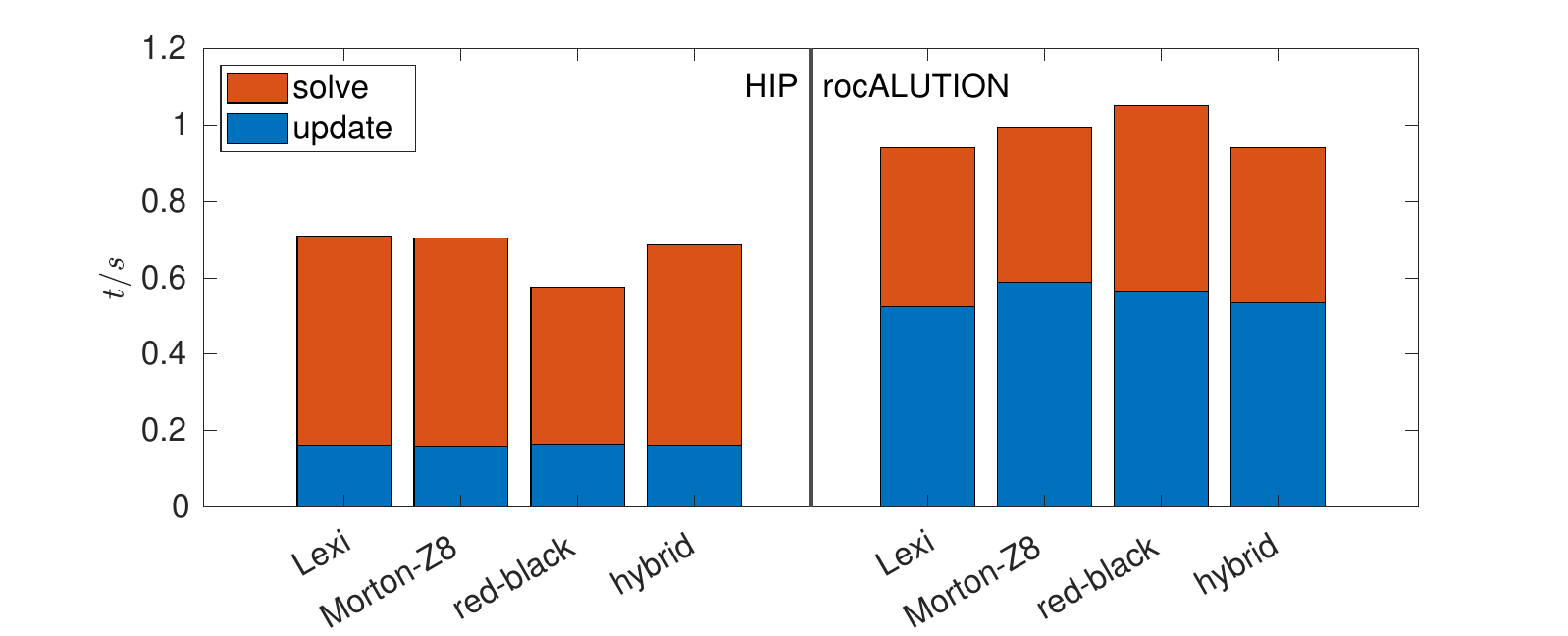}
    \caption{Execution times for the large-scale problem with Neumann boundary conditions using different mesh orderings. Results are shown for the HIP backend (left half) and the \texttt{rocALUTION} backend (right half). The hybrid ordering combines a Morton Z ordering of block size 8 on the finest mesh level with a red-black ordering on the coarser levels.}
    \label{fig:viper_reorder}
\end{figure}

We evaluate the impact of different mesh orderings on the HIP and \texttt{rocALUTION} backends in Fig.~\ref{fig:viper_reorder}, using the large-scale problem with Neumann boundary conditions executed on the VIPER-GPU system. While the lexicographic and Morton orderings yield comparable performance, we observe a significant $25\%$ speedup in the solve phase when using red-black ordering. Since in \texttt{GRILLIX} and \texttt{GENE-X} the stencils from other terms may be adversely affected by red-black ordering compared to Morton ordering, we also investigate a hybrid approach. In this setup, the finest mesh level uses Morton ordering, while coarser levels, relevant only to the multigrid preconditioner, use red-black ordering. This hybrid strategy results in a minor $5\%$ speedup in the solve phase, reflecting the fact that the most computationally intensive operations occur on the finest level, as discussed in section \ref{sec:performance_breakdown}. With \texttt{rocALUTION}, we observe performance variations of up to $15\%$ depending on the mesh ordering, but here red-black ordering yields the poorest results.

\subsection{Integrated test case in \texttt{GRILLIX}}
The final test evaluates the full integration of the GPU solver backends within a production simulation of \texttt{GRILLIX}. \texttt{GRILLIX} solves the global electromagnetic drift-reduced Braginskii fluid turbulence model, with extensions for low-collisionality effects. To model neutral gas dynamics and their interaction with the plasma, a three-moment neutrals fluid model is solved additionally. For detailed information on the physical model and the governing equations in \texttt{GRILLIX}, we refer to \cite{zholobenko:hmode24}. We use the simulation described there, which employs 16 poloidal planes, to verify and benchmark the GPU backends. Specifically, we resumed the simulation from the last available state, executed 40 additional time steps, and performed performance profiling over the final 20 steps.

We performed the benchmark on the VIPER-GPU@MPCDF system, where each node features 2 AMD Instinct MI300A APUs. Each of the 16 poloidal planes runs on a separate MPI process, which is assigned to a dedicated APU, comprising 24 CPU cores used for OpenMP parallelisation and a GPU. MPI communication is only required for evaluating FCI operations \--- such as computing parallel gradients along magnetic field lines \--- where the stencil spans adjacent poloidal planes. The field solver operates independently and is unaware of MPI, with the field equation on each plane solved on the GPU associated with the corresponding MPI process.

\begin{figure}
    \includegraphics[trim=0 0 0 0, clip=true, width=0.5\textwidth]{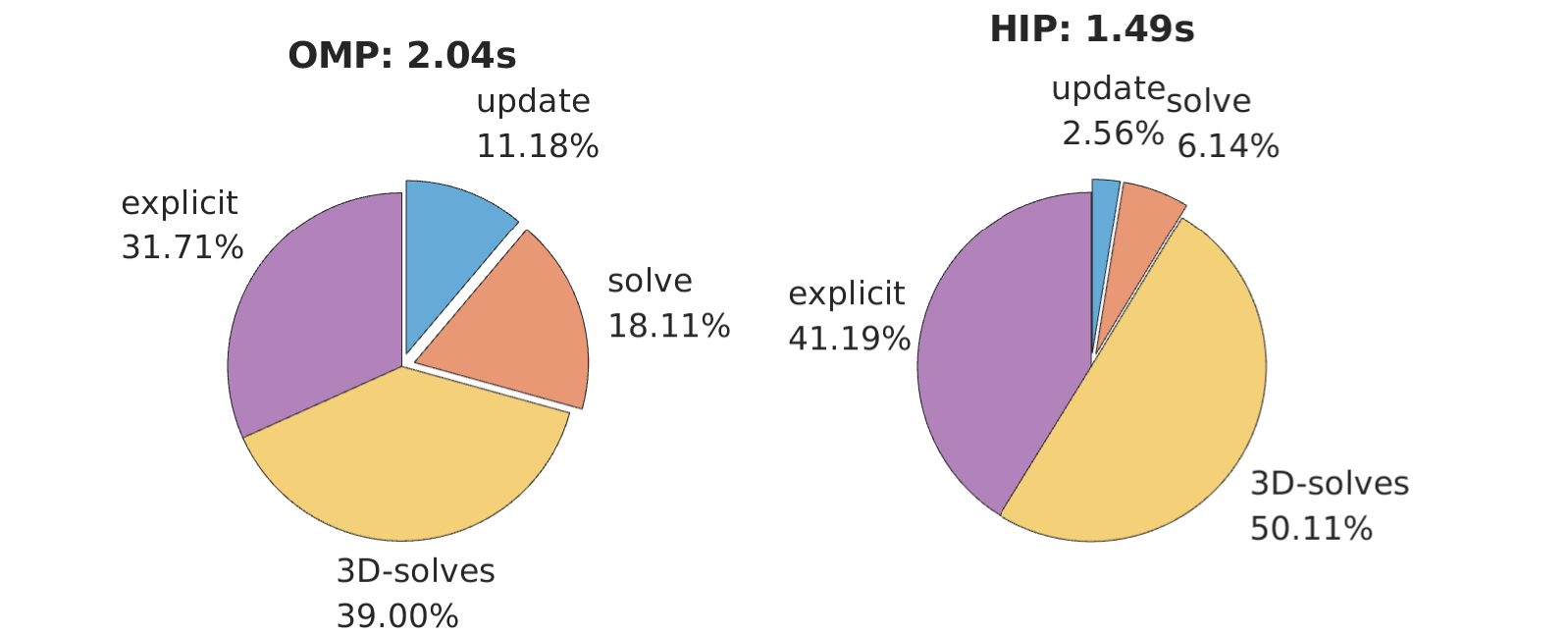}
    \caption{Breakdown of computational time spent in different sections of \texttt{GRILLIX} using the OMP backend (left) and the HIP backend (right) for the field solver. The explicit and 3D-solves parts are still not offloaded to GPUs. The average time per time step decreases from 2.04 s with the OMP backend to 1.49 s with the HIP backend.}
    \label{fig:tstep_pie}
\end{figure}

Figure~\ref{fig:tstep_pie} shows a breakdown of the computational time spent in different sections of a time step, comparing the OpenMP and HIP backends for the field solver. The computational workload can roughly be divided into three main components. First, various field solves (update + solve) are performed to compute the electrostatic potential and the parallel perturbed part of the electromagnetic potential. Additional field solves also implicitly advance stiff diffusion terms in the neutrals model acting within poloidal planes \cite{eder:neutrals25}. Second, the computation of explicit terms is carried out through OpenMP-parallelised loops. Third, three-dimensional solves are used to advance stiff terms associated with parallel heat conduction and parallel momentum damping implicit in time. These 3D-solves employ iterative methods \cite{cunha:pim95} that act across all poloidal planes and involve FCI matrices. 

With the OpenMP backend for the field solver, all three components have similar computational costs. When switching to the GPU backend, the field solve is accelerated by approximately a factor of $5\times$, reducing its share of the total runtime from $29\%$ to $9\%$. Meanwhile, the computation times for the other components remain roughly constant, and their share is increased, as they have not yet been offloaded to the GPU.

\section{Conclusions and outlook}
\label{sec:Conclusions_and_outlook}
Field solvers play a pivotal role in edge/SOL turbulence codes. Unlike in core turbulence codes, the need for a full-f description in edge/SOL simulations leads to a generalised form of elliptic equations, where the elliptic operator becomes time-dependent due to polarization. Additional complexity arises from the complex geometry of diverted devices, including challenging boundary shapes. 

We propose a finite difference discretization scheme applied to a set of independent poloidal planes. Each plane is locally Cartesian yet logically unstructured, enabling the mesh to approximately conform to complex boundary geometries. The time dependence of the resulting discrete matrix motivates the use of an iterative solver. We employ a flexible GMRES solver equipped with a geometric multigrid preconditioner. Multigrid methods are particularly well-suited for elliptic problems, offering convergence rates that do not deteriorate with mesh refinement. The field solver is a major part of the \texttt{PARALLAX} library, which forms the foundation of the fluid edge turbulence codes \texttt{GRILLIX} and \texttt{GENE-X}, which adopts a gyrokinetic model. The solver follows an abstract design pattern, facilitating seamless integration with different solvers and backends, including external library solutions. Due to the limitations of Fortran in GPU porting, the \texttt{PAccX} library was developed, dedicated to a C++ implementation of the field solver with support for both CUDA and HIP backends. Additionally, \texttt{PAccX} offers a wrapper to the third-party \texttt{rocALUTION} library as an alternate solver backend.

We verified both the expected discretization order for the elliptic equation and the linear scaling of the solver’s computational cost with problem size, confirming the expected $O(N)$ behavior. The GPU backends provide a significant performance boost, accelerating the solver by a factor of approximately $4\times$ for smaller problem sizes and up to $15\times$ for larger ones, relative to the OpenMP-parallelised CPU implementation. This performance gain is consistent across different hardware architectures using both the CUDA and HIP backends. Compared to \texttt{rocALUTION}, our implementation is $2\times$ to $4\times$ faster, and has demonstrated robustness through sustained use in \texttt{GRILLIX} and \texttt{GENE-X}. Further performance improvements may be achieved through fine-tuning, as was exemplified by the mesh reordering. Future efforts will focus on optimizing the direct solver at the coarsest level and potentially offering BiCGStab as an alternative to the GMRES solver.

The integration of the GPU-accelerated solver into a full \texttt{GRILLIX} simulation has been successfully verified. While the expected speedup is achieved in the components involving the solver, the overall performance gain is currently limited to approximately $25\%$, as other parts of the code \--- still running on the CPUs \--- remain a bottleneck. Nevertheless, this represents a significant milestone: the field solver is the algorithmically most complex component, and offloading the remaining parts \--- such as explicit terms and 3D solves, which primarily involve parallelisation of simple loops \--- are comparatively straightforward. Separately, the GPU offload of \texttt{GENE-X} is being actively pursued \cite{Trilaksono2025}.

\texttt{PARALLAX} and \texttt{PAccX} are publicly available as part of the \texttt{phoenix-public} Git group, hosted by MPCDF at:
\url{https://gitlab.mpcdf.mpg.de/phoenix-public}.

\section*{Code and data availability}
The source codes for \texttt{PARALLAX} and \texttt{PAccX} are publicly available at \url{https://gitlab.mpcdf.mpg.de/phoenix-public/parallax.git} and \url{https://gitlab.mpcdf.mpg.de/phoenix-public/paccx.git}, respectively. The input and output files for the benchmarks and verifications are archived at the Max Planck Computing and Data Facility and will be made available upon reasonable request.

\section*{CRediT authorship contribution statement}
\noindent
\textbf{Andreas Stegmeir:} Conceptualization, Formal analysis, Investigation, Methodology, Project administration, Software, Supervision, Validation, Writing \-- original draft.
\textbf{Cristian Lalescu:} Conceptualization, Formal analysis, Investigation, Methodology, Software, Validation.
\textbf{Mou Lin:} Formal analysis, Investigation, Methodology, Software, Validation
\textbf{Jordy Trilaksono:} Methodology, Software, Validation.
\textbf{Nicola Varini:} Methodology, Software, Validation.
\textbf{Tilman Dannert}: Conceptualization, Project administration, Supervision.

\section*{Declaration of generative AI and AI-assisted technologies in the writing process}

During the preparation of this work the authors used OpenAI's ChatGPT service in order to refine the language and grammar of the manuscript. After using this service, the authors reviewed and edited the content as needed and take full responsibility for the content of the published article.

\section*{Acknowledgements}
This work has been carried out within the framework of the EUROfusion Consortium, funded by the European Union via the Euratom Research and Training Programme (Grant Agreement No 101052200 – EUROfusion). 
Views and opinions expressed are those of the author(s) only and do not necessarily reflect those of the European Union or the European Commission. Neither the European Union nor the European Commission can be held responsible for them.




\end{document}